\documentclass[onecolumn, a4paper, oneside, onecolumn, titlepage]{article}

\usepackage{lineno}					
\usepackage{amsmath}				
\usepackage{amsfonts}				
\usepackage{setspace}				
\usepackage[]{natbib}               
\usepackage[nocaption]{optional}	
\usepackage{subfig}					
\usepackage{graphicx}				
\usepackage{booktabs}				
\usepackage{siunitx}				

\usepackage[%
	lmargin=2.5cm,%
	rmargin=2.5cm,%
	tmargin=3cm,%
	bmargin=3cm%
]{geometry}%

\newcommand*\patchenviroforlineno[1]{%
	\expandafter\let\csname old#1\expandafter\endcsname\csname #1\endcsname %
	\expandafter\let\csname oldend#1\expandafter\endcsname\csname end#1\endcsname %
	\renewenvironment{#1}%
		{\linenomath\csname old#1\endcsname}%
		{\csname oldend#1\endcsname\endlinenomath}%
}%

\newcommand*\patchbothenviroforlineno[1]{%
	\patchenviroforlineno{#1}%
	\patchenviroforlineno{#1*}
}%

\AtBeginDocument{%
	\patchbothenviroforlineno{equation}%
	\patchbothenviroforlineno{align}%
	\patchbothenviroforlineno{flalign}%
	\patchbothenviroforlineno{alignat}%
	\patchbothenviroforlineno{gather}%
	\patchbothenviroforlineno{multline}%
}%

\begin{document}
\begin{titlepage}
	\vspace{\stretch{1}}
	\begin{center}
		{ \huge \bfseries %
		Causes and consequences of dispersal in biodiverse spatially structured systems: what is old and what is new?
		} \\ %
		\vspace{\stretch{1}}
		Emanuel A. Fronhofer$^{1, *}$, Dries Bonte$^2$, Elvire Bestion$^3$, Julien Cote$^4$, Jhelam N. Deshpande$^1$, Alison B. Duncan$^1$, Thomas Hovestadt$^5$, Oliver Kaltz$^1$, Sally Keith$^6$, Hanna Kokko$^7$, Delphine Legrand$^3$, Sarthak P. Malusare$^1$, Thomas Parmentier$^{2, 8}$, Camille Saade$^1$, Nicolas Schtickzelle$^9$, Giacomo Zilio$^1$ and François Massol$^{10}$
	\end{center}
	\vspace{\stretch{0.25}}
	\begin{enumerate}
	\item ISEM, University of Montpellier, CNRS, IRD, EPHE, Montpellier, France
	\item Terrestrial Ecology Unit (TEREC), Department of Biology, Ghent University, K.L. Ledeganckstraat 35, B-9000 Ghent, Belgium
	\item Station d’Ecologie Théorique et Expérimentale, CNRS, UAR 2029, F-09200 Moulis, France
	\item Laboratoire Évolution \& Diversité Biologique, CNRS, Université Toulouse III Paul Sabatier, IRD; UMR5174, 118 route de Narbonne, F-31062 Toulouse, France
	\item Department Animal Ecology and Tropical Biology, Biozentrum, University of Würzburg, Würzburg, Germany
	\item Lancaster Environment Centre, Lancaster University, Lancaster, LA1 4YQ, UK
	\item Institute of Organismic and Molecular Evolution, Johannes Gutenberg University, 55128 Mainz, Germany
	\item Research Unit of Environmental and Evolutionary Biology, Namur Institute of Complex Systems, and Institute of Life, Earth, and the Environment, University of Namur, Rue de Bruxelles 61, 5000 Namur, Belgium
	\item Earth and Life Institute, UCLouvain, B-1348 Louvain-la-Neuve, Belgium
	\item Institut Pasteur de Lille, Univ. Lille, CNRS, Inserm, CHU Lille, U1019 - UMR 9017 - CIIL - Center for Infection and Immunity of Lille, Lille, France
	\end{enumerate}
	\vspace{\stretch{2}}
	\begin{flushright}
		\textbf{Correspondence Details}\\
		Emanuel A. Fronhofer\\
		Institut des Sciences de l'Evolution de Montpellier, UMR5554\\
		Universit\'e de Montpellier, CC065, Place E. Bataillon, 34095 Montpellier Cedex 5, France\\
		phone: +33 (0) 4 67 14 31 82\\
		email: emanuel.fronhofer@umontpellier.fr
	\end{flushright}
\end{titlepage}

\doublespacing
\begin{abstract}
Dispersal is a well recognized driver of ecological and evolutionary dynamics, and simultaneously an evolving trait. Dispersal evolution has traditionally been studied in single-species metapopulations so that it remains unclear how dispersal evolves in spatially structured communities and food webs. Since most natural systems are biodiverse and spatially structured, and thus affected by dispersal and its evolution, this knowledge gap should be bridged.

Here we discuss whether knowledge established in single-species systems holds in spatially structured multispecies systems and highlight generally valid and fundamental principles. Most biotic interactions form the ecological theatre for the evolutionary dispersal play because interactions mediate patterns of fitness expectations in space and time. While this allows for a simple transposition of certain known drivers to a multispecies context, other drivers may require more complex transpositions, or might not be transferred. We discuss an important quantitative modulator of dispersal evolution in the increased trait dimensionality of biodiverse meta-systems and an additional driver in co-dispersal.

We speculate that scale and selection pressure mismatches due to co-dispersal, together with increased trait dimensionality may lead to slower and more ``diffuse'' evolution in biodiverse meta-systems. Open questions and potential consequences in both ecological and evolutionary terms call for more investigation.
\end{abstract}

\newpage
\section*{Introduction}
Dispersal is a central life-history trait \citep{Bonte2017}, defined as the process by which an organism reproduces away from where it is born \citep{Clobert2009, Bowler2005, Ronce2007, Clobert2012, Duputie2013}. By definition, dispersal is therefore different from seasonal migration or foraging \citep{Schlaegel2020}, which do not lead to gene flow. Dispersal can be quantified in multiple ways, such as through dispersal distance, dispersal frequency in long-lived mobile organisms, the proportion of dispersed offspring, or even the proportion of immigrants among a population’s reproductive cohort. Dispersal affects many facets of population biology, such as population densities, phenotypic and genotypic composition, species ranges and spatial distributions, their persistence in the face of perturbations, or their ability to adapt to local conditions \citep{Sabelis2005, Grainger2016, Bonte2017, Massol2017, Fronhofer2017}. Ultimate and proximate determinants of dispersal have received substantial interest from both empiricists and theoreticians and are integrated in the framework of evolutionary conservation biology \citep{Olivieri2016}. Dispersal also plays a pivotal role at the interface between ecology and evolution and can drive eco-evolutionary feedbacks \citep{Govaert2019, Fronhofer2023}.

As an important driver of ecological properties across scales that encompass individuals, populations, communities and ecosystems, dispersal is under a wide range of selective pressures which we discuss in detail below \citep[see also][]{Duputie2013}. Most of our understanding of these processes comes from single-species studies, or theoretical work that considers selection acting within one species. This leaves us with a dearth of studies that have tackled the evolutionary determinants of dispersal in a community or food web context \citep[but see e.g.,][; for a review focused on range dynamics see \citealt{Kubisch2014}]{Kinlan2003, Baiser2013}.

While intra-specific interactions and their effects on dispersal have been studied extensively \citep{Hamilton1977, Leturque2002, Ajar2003, Massol2011, Kisdi2016}, dispersal evolution models that do include inter-specific interactions (e.g., predation, facilitation, parasitism) traditionally consider these interactions only implicitly or in a very simplified manner, such as seed predation accounting for the extra mortality of dispersed and dormant seeds \citep{Vitalis2013}. The lack of a community angle is perhaps not surprising given that the topic of dispersal has been historically entwined with population biology, demography, life-history evolution and the metapopulation framework \citep{Hanski1997}. Yet, this oversight constitutes a substantial knowledge gap since interactions between organisms extend beyond intraspecific competition for limiting resources within one species, and encompass a continuum of antagonistic and mutualistic interactions with various degrees of immediacy, from event-like (e.g., predation) to lifelong interactions (obligate symbiotic relationships). All interactions are likely to affect selection on, and the consequences of dispersal. For instance, a large proportion of parasitic and mutualistic symbionts disperse thanks to the movement of their hosts \citep[e.g.][]{Pagan2022, Matthews2023, Parmentier2021} and plants often make use of animal vectors to disperse their seeds \citep[e.g.,][]{Heleno2013}.

We address this knowledge gap in two parts. First, we set the scene by synthesizing what is already known about dispersal evolution from the single-species perspective. Second, we discuss how far these known mechanisms can be transposed to a multi-species context and highlight novel aspects that may be relevant in driving dispersal in multispecies meta-systems. In both cases, we tackle the ultimate causes of dispersal, i.e., the selective pressures on dispersal, as well as the proximate causes of dispersal, which include the factors driving dispersal plasticity. We end by discussing new perspectives relevant to a multispecies world and future research directions.

\section*{A synthesis on dispersal in a single-species world}
\subsection*{Why disperse?}
Comprehensive reviews on dispersal evolution have listed the different selective pressures on dispersal and how their combination results in positive, negative, disruptive or stabilizing selection \citep{Bowler2005, Ronce2007, Duputie2013}. In brief, dispersal is selected for in situations in which there is temporal variability in environmental conditions \citep{Gadgil1971}, perturbations \citep{Comins1980}, kin competition \citep{Hamilton1977}, and inbreeding depression \citep{Bengtsson1978}. In contrast, stable spatial heterogeneity \citep{Balkau1973, Hastings1983} and direct dispersal costs \citep{Bonte2012} favour non-dispersing individuals. 

Beyond these basic selective forces, the demographic dynamics of populations play a complex role in dispersal evolution. For example, chaotic population dynamics can select for higher dispersal rates \citep{Holt1996}, while Allee effects can select against dispersal \citep{Travis2002}. When local dynamics are accounted for, increasing population extinction rate does not always select for higher dispersal rates, because frequent local extinctions tend to weaken the overall intensity of competition \citep[][but see \citealt{Poethke2003}]{Ronce2000a}.

Interestingly, the timing of dispersal, and the dependency on specific life stages, has a major impact on both the evolution of dispersal rates and distances \citep[disruptive vs. stabilizing;][]{Massol2015}. Combining selective pressures on dispersal does not always lead to ``averaging influences'' and can in fact induce disruptive selection on dispersal. For instance, a combination of spatial heterogeneity, kin competition and dispersal costs can lead to the evolutionary branching of dispersal rates \citep{Massol2011, Kisdi2016, Parvinen2020}.

The evolution of dispersal can interact with evolutionary changes in other traits. Some of the best studied traits in this context include traits linked to local adaptation \citep{Kisdi2002, Billiard2005, Berdahl2015, Jacob2017} and self-fertilization \citep{Cheptou2009, Iritani2017, Rodger2018}, and the rationale behind this interaction is often very similar \citep[see][for a population genetics model which considers dispersal evolution in response to local adaptation and inbreeding depression]{Wiener1993}. The joint evolution of dispersal with local adaptation can result in disruptive selection or alternative evolutionarily stable strategies in temporally variable environments. Indeed, habitat generalists perceive and/or react to this variability to a lesser degree than specialists, hence associating selective pressures for generalism with selective pressures against dispersal \citep[][see also \citealt{Massol2011b}, considering selfing individuals as ``pollination generalists'']{Kisdi2002}. In spatially heterogeneous habitats, generalists tend to disperse more than specialists \citep[][see also \citealt{Rodger2018} for the selfing evolution equivalent]{Berdahl2015}. 

The evolution of dispersal can thus be affected by the evolution of another trait through reciprocal selective pressures (i.e., joint evolution; as a response to fitness effects imposed by the evolution of another trait, or conversely, other traits evolve in response to the evolution of dispersal). This naturally generates correlations between dispersal and other traits, i.e. ``dispersal syndromes''. However, this is but one of the reasons behind the existence of dispersal syndromes \citep{Ronce2012}, as syndromes could also emerge from allocation trade-offs (e.g., foraging vs. dispersal), genetic correlations and pleiotropy, structural constraints (e.g., allometric scaling), or shared selective pressures (e.g., inbreeding depression affecting both the evolution of selfing and dispersal when accounting for inbreeding between relatives).

While the evolution of dispersal has traditionally been considered in an equilibrium metapopulation, dispersal can also evolve in non-equilibrium contexts. In particular, considering the evolutionary dynamics of populations along an invasion front has highlighted the existence of ``spatial selection'', by which higher dispersal rates are selected at the front of an advancing invasion wave \citep{Phillips2010a, Perkins2013, Hanski2011}. This can stem from a combination of factors: A higher probability to end up at the front of the wave with higher dispersal rate \citep{Shine2011}. This in turn can trigger passive assortative mating in sexually reproducing species, and a gradient of local competition strength since denser populations far from the front harbour more intense competition, which can affect dispersal evolution via trade-offs between competitiveness and dispersal \citep{Burton2010}. Alternatively, a general trade-off between foraging and dispersal can also generate the same selection for higher dispersal rates at the front \citep{Fronhofer2015b}. Spatial selection is interesting as it can lead to accelerating range expansions which may explain the rapid recolonization of trees in temperate latitudes since the last glacial period \citep[also known as Reid’s paradox; see][]{Phillips2008}. The evolution of higher dispersal rates at the invasion front is also reminiscent of the evolution of lower dispersal in ageing demes \citep{Olivieri1995} --- in both cases, newly colonised patches harbour individuals with higher dispersal rates than older ones.

\subsection*{Dispersal plasticity}
Dispersal need not be a fixed trait, but can respond plastically to internal and external cues \citep{Clobert2009}. The biotic and abiotic contexts can affect both emigration and immigration \citep[see e.g.][]{Weigang2017}. The extrinsic context typically includes patch-quality proxies \citep[or simply patch types, e.g.][]{McPeek1992}, such as patch size \citep{Gros2006}, patch age \citep{Ronce2005}, population growth rate, population density \citep{Poethke2007, Parvinen2012, Fronhofer2017b}, carrying capacity \citep{Poethke2002}, as well as available information about such proxies \citep[number of mated or reproducing conspecifics, e.g. ][]{Doligez2003}. Information about the matrix habitat can also modulate context-dependent dispersal decisions \citep{Stamps1987, Moilanen1998}. The reliability of cues is an important parameter affecting the evolution of dispersal plasticity \citep{McNamara2011}.

Intrinsic individual-level conditions affecting dispersal include sex \citep{Hardouin2012, Hovestadt2014, Li2019}, age \citep{Ronce2000b}, developmental stage \citep{Cayuela2020}, body size or condition \citep{Bonte2009a, Baines2015}, or any proxy of competitiveness, for example. Intrinsic and extrinsic cues can of course be used at the same time and interact to determine dispersal decisions \citep[e.g.][]{Deshpandeinprep.a}.

\subsection*{Consequences of dispersal}
In a single-species context, dispersal is known to have multiple consequences, both ecologically and evolutionarily \citep{Bowler2005, Peniston2023}. From an evolutionary point of view, dispersal promotes genetic mixing: immigrants bring new alleles into populations and thus increase local genetic diversity. Conversely, a (very) low dispersal rate between populations increases differentiation. The mixing effect of dispersal thereby constrains local adaptation and may lead to ``migration load'' known from classic population genetics models. This is especially striking in continuous space models of adaptation on environmental gradients, where high dispersal rates tend to impede local adaptation \citep{Pease1989, Kirkpatrick1997, AlleaumeBenharira2005, Duputie2012}. The stage at which dispersal takes place is also critical to the evolution of local adaptation, so that higher dispersal rates might actually not affect local adaptation in certain cases \citep{Ravigne2006a, Ravigne2009, Debarre2011, Massol2013}.

From an ecological point of view, dispersal is also an important parameter controlling the dynamics of metapopulations in at least two different ways. First, dispersal drives the synchrony of local population dynamics, but in a non-linear manner that also depends on other factors such as perturbation frequency, the relative speed of population dynamics of the focal species vs. its natural enemies or resources, or the spatial extent of dispersal \citep{Doebeli1995, Lande1999, Ylikarjula2000, Koelle2005, Goldwyn2008, Oliver2017}. Second, dispersal determines the resilience of metapopulations to perturbations as higher dispersal rates can lead to more rapid recolonisation after perturbation. 

At the metapopulation level, an important effect of higher dispersal rates is higher occupancy, as long as dispersal translates into higher colonization rates \citep{Hanski1993, Etienne2002, Jansen2007}. In other words, species occupy more patches when they disperse more. However, this general rule can be counteracted in two important ways. First, when the costs of dispersal are substantial, dispersal rates that are too high can induce a decrease in colonization rate since emigrating effectively leads to a decrease in local birth rate \citep{Jansen2007}. Second, when local populations experience an Allee effect, dispersal hinders the maintenance of population sizes above the Allee threshold \citep{Travis2002}.

Overall it is important to note that these general effects are often based on random, unconditional dispersal. They may therefore be counteracted or even inverted by conditional emigration and habitat selection \citep{Jacob2017, Mortier2019}.

\section*{A more ``complex'' world --- dispersal in communities, food webs and ecosystems}
The central question we raise here is: What baseline assumptions regarding the determination of dispersal --- both at the ultimate and proximate levels --- are changed when interspecific interactions are taken into account? This question is currently unresolved, both empirically and theoretically. For the sake of simplicity we will first discuss potential impacts in horizontal, competitive communities before moving on to more complex interaction webs.

\subsection*{Horizontal interactions --- metacommunities}
In metacommunities \citep{Leibold2004, Thompson2020} with horizontal interactions (i.e., non-trophic interactions), the evolution of dispersal has been considered in the context of competition-colonization trade-offs \citep{Laroche2016, Cai2022}. The parallels to single-species work and thinking are unmistakable, and these parallels also become obvious when considering dispersal plasticity, especially density-dependent dispersal in a metacommunity context, where single-species theory can directly be transposed to the multispecies context taking into account differential strengths of competition \citep{Fronhofer2015c}.

While the parallels between single- and multi-species drivers of evolution are clear when assuming asexual reproduction since genotypes are then indistinguishable from species, sexual reproduction and kinship may a priori prevent us from extrapolating from single to multispecies meta-systems. However, this added complexity may rather be a matter of quantitative effects than qualitative differences.

For instance, recombination of genetic material may happen at the interspecific level via hybridisation which is an important eco-evolutionary mechanism \citep[][]{Oziolor2019} and an analogue to sexual reproduction. Inbreeding depression affects single-species dispersal evolution, but it has no clear multispecies equivalent, except maybe if we assume the unlikely case of hybrid vigour, playing a role that is similar to overdominance in single-species models. Similarly, kin competition seems irrelevant at the interspecific level as an ultimate determinant of dispersal, since the probability of sharing the same allele between individuals from different species should, a priori, be very low. However, this may not be the case due to interspecific gene flow from hybridization, as discussed above, or due to incomplete lineage sorting \citep[ancestral polymorphism between species; see e.g.,][]{Klein1998, Jamie2020}, for example. When considering proximate determinants of dispersal, phylogenetic proximity of interacting species might represent an analogue cue to relatedness in single-species cases since close species might be more likely to have similar niches, and hence strongly compete for resources \citep[see][for a more detailed discussion on the link between phylogenetic proximity and traits]{Mayfield2010}.

While these examples are speculative and call for more formal investigations, they highlight that, at least for horizontal communities, the classic mechanisms discussed in single-species models mostly hold, however, likely with altered relative importances.

\subsection*{Vertical interactions --- meta-food webs}
Does the above conclusion hold when moving to vertical interactions such as predation and parasitism? Existing theory can inform us here. For instance, predation affects the evolution of dispersal in predators and preys alike since it intrinsically controls both species dynamics and thus the spatio-temporal dynamics in (expected) inclusive fitness \citep{Chaianunporn2012, Chaianunporn2019, Drown2013, Travis2013a, Hochberg1999, Price2009}. This patterning of spatial fitness expectations, that here relies on predator-prey cycles, is different from what can be expected in a simple horizontal competitive setting or even under mutualism \citep{Chaianunporn2012} and represents an important driver of dispersal evolution. Note that, as for horizontal communities, the fundamental forces driving dispersal evolution do not differ from those in single species discussed above \citep{Poethke2010, Deshpande2021}. This example shows how, in \citet{Hutchinson1965}'s words, interspecific interactions can form the ecological theatre in which the evolutionary play of the focal species unfolds \citep{Fronhofer2023}. This biotic determination of selection pressures is interesting since it may actually reduce ``parameter space'': while theoretically any spatio-temporal autocorrelation of fitness expectations can be assumed, biotic interactions, by generating characteristic population dynamics in interacting species, here determine, and constrain, the nature of this autocorrelation. Importantly, we have models that allow us to capture the essence of these patterns \citep[e.g.][for a discrete-time discrete-fitness expectation model]{Massol2015}.

Going one step further, the study of the joint evolution of interaction strength or selectivity and dispersal has almost never been undertaken. This is of interest because, the strength of interspecific interactions determines the extent to which interacting partner densities impact each other, hence the drivers of dispersal evolution \citep{Chaianunporn2012, Chaianunporn2019}. Further, via its feedback on the demography of interacting species, dispersal can impact the evolution of interaction strength. Thus, the joint evolution of dispersal and interaction strength is likely to further constrain population dynamics of interacting partners and dispersal evolution. \citet{Calcagno2023} modelled the evolution of colonization rates in a multi-trophic occupancy setting in which colonisation rate trades off with competitive ability, but did not incorporate changes in top-down or bottom-up controls as consequences of colonization evolution \citep[see also ][for the two-level, predator-prey case]{Pillai2012}. In the case of symbiotic interactions, \citet{Ledru2022} tackled the question of dispersal evolution together with that of the evolution of mutualism. Their conclusions that localized competition favours the evolution of mutualism and that mutualistic types tend to be associated with lower dispersal rates converge with studies on the joint evolution of intraspecific altruism and dispersal \citep{LeGalliard2005, Mullon2017}.

\subsection*{Biotic proximate determinants of dispersal: co-dispersal and mobile links}
As discussed above, proximate biotic drivers of dispersal that lead to context-dependency are clearly present at the interspecific level \citep{Fronhofer2018, Cote2022} but they can overall be explained relying on theory developed for single-specific systems.

However, an additional process that is not addressed by classic studies is the occurrence of ``co-dispersal'', i.e., cases where the two interacting partners disperse together. This echoes the idea of mobile links \citep{Lundberg2003} --- organisms connecting habitats via transport of resources or processes (think of hippopotami linking terrestrial and aquatic ecosystems via transfer of matter; for many more examples see \citealt{Massol2011a} and \citealt{Gounand2018a}). Here, these mobile links transport other organisms and their genes. 

For many symbionts, but also for animal-dispersed plant seeds or phoretically dispersed organisms \citep[organisms travelling on other organisms, e.g.,][]{Fronhofer2023}, the movements of the host or vector may be unrelated to dispersal (being, instead, related to foraging or seasonal migration), and yet result in dispersal for the symbiont species. For example, foraging pollinators may disperse parasites among plant species \citep{Graystock2015} or female mosquitoes, blood feeding, may vector different parasites (e.g., viruses or malaria). In such a situation the usual dispersal drivers do not apply, and are replaced by the drivers of movement for the host or vector \citep{Calcagno2014, RiotteLambert2020}. This mismatch in spatial scale may lead to conflicting selection pressures.

Except in the case of obligate mutualisms, this dependence on dispersal is almost always asymmetric, e.g., plants need animals for seed dispersal, but animals can disperse without seeds; symbionts often need their host to disperse, but symbiont-free hosts can disperse. In these cases, the dispersal of the host organism may be controlled by both its own genes and those of the interacting species, through its extended phenotype. For instance, symbionts might manipulate the dispersal of their hosts, so that infected individuals disperse more or less than uninfected ones \citep{Goodacre2009, Fellous2011, Zilio2021}. Conversely, infection could trigger an adaptive state-dependent dispersal strategy in the host \citep{Debeffe2014, Iritani2014, Deshpande2021}. Specifically, endosymbionts and in particular intracellular endosymbionts, on the one hand, will co-disperse since they are an ``inherent part'' of the host, unless disposed of by its immune system. There is evidence that endosymbionts might induce dispersal, select for dispersal in their host, bias immigration success, or affect dispersal distance (e.g., Wolbachia, Rickettsia: \citealt{Bonte2009} and \citealt{Goodacre2009}; microbiome and metacommunity: \citealt{Brown2019}). Ectosymbionts, on the other hand, will not exhibit co-dispersal fidelity with certainty, but can induce dispersal (and co-dispersal). However, the processes and dynamics are different in ectosymbionts since they may be affected by the environment surrounding their host and hosts may attempt to modify symbiotic load by relocating to a different habitat \citep{Brown1992a}. 

While manipulation by the symbiont represents one way of addressing the above-mentioned conflict over selection pressures, adaptive dispersal strategies for the host may also lead to curing \citep[i.e., extinction of the parasite; ][]{Bonte2009}. To make the issue more complex, hosts can be co-infected by multiple parasites where competition between parasites can impact transmission (dispersal). Finally, parasites may be co-transmitted \citep{Syller2011}, whereby one relies on another for transmission to new hosts. These complexities are beyond the scope of the current paper.

The consequences of co-dispersal can be many fold. From an evolutionary point of view, dispersing with or without a symbiont is known to be different, and can lead to different dispersal and trait evolutionary trajectories \citep{SchlippeJusticia2022, Zilio2023}. These effects can occur under both mutualistic and parasitic symbioses \citep{Martignoni2023}, while parasitic symbioses are also prone to evolve towards manipulation of host dispersal \citep{Mayer2021, Noergaard2021}. Unfortunately, similarly detailed analysis involving more realistic meta-food webs are missing. In conclusion, while we above highlight the general validity of dispersal drivers identified at the intra-specific level, co-dispersal clearly represents an additional driver at the inter-specific level.

\subsection*{Increased trait dimensionality in metacommunities and meta-food webs}
In general, moving from a single- to a multi-species context should increase system complexity and thereby also the dimensionality of the relevant trait space. \citet{Debarre2014} show that increased trait dimensionality tends to destabilise evolutionary equilibria. This in turn increases opportunities for evolutionary branching and can therefore favour diversification, which could lead to polymorphisms of dispersal or other traits \citep[see also][]{Doebeli2010} with (quasi-)equilibrium reached when the systems settles --- across different species --- in an ideal-free fitness distribution \citep{Chaianunporn2011}. By contrast, a more complex biotic environment can be understood as inducing a ``cost of complexity'' \citep{Orr2000}, i.e., adaptation is slower when it is required on more dimensions, or, if mutations are more pleiotropic (and thus quicker), phenotypic variation can decrease with the number of dimensions \citep[the ``preservation of perfection'' effect,][]{Waxman1998}.

From  \citet{Debarre2014} we may predict that diversity in species interactions begets diversity (at least in trait distributions; see also \citealt{Calcagno2017} outside the context of dispersal evolution). At the same time, the cost of complexity \citep{Orr2000} and the preservation of perfection \citep{Waxman1998} may imply that evolution will be slower in more complex systems. This has interesting consequences for eco-evolutionary feedbacks, since complexity may therefore prevent eco-evolutionary feedbacks by decoupling ecological and evolutionary timescales \citep[see also][]{Fronhofer2023}. The ultimate consequence of this interplay may be that dispersal evolution is overall slower and more ``diffuse'' in multispecies meta-systems. Potentially, this could lead to convergence in dispersal strategies across species or neutral coexistence of diversity. One can speculate that, as a consequence, dispersal limitation could be less likely overcome by evolution which implies that current models which are largely single species models would drastically overestimate the potential for adaptation and range shifts.

\section*{Discussion}
Despite a strong interest in dispersal, its effects on evolutionary dynamics and many ecological properties, ecologists and evolutionary biologists have yet to properly tackle the determinants of dispersal in community and food web contexts. This integration is especially relevant since dispersal has important consequences at the multispecies level. These consequences include effects on biodiversity and its geographic distribution \citep{Leibold2004, Gilman2010, Kubisch2014}, but also effects on system complexity \citep{Pillai2011}, stability \citep{Gravel2016a}, coexistence and functioning \citep{Gounand2018a}. For instance, alpha and beta diversity patterns are drastically influenced by dispersal \citep{Leibold2004, Economo2008, Carrara2012, Saade2022} as well as by evolutionary dynamics under global change \citep{Norberg2012, Thompson2019, Fronhofer2023}. Yet, how these patterns are affected by dispersal evolution within the metacommunity and meta-food web context remains unclear. The same is true for our understanding of how dispersal evolution affects community resilience and assembly \citep[for a detailed discussion see][]{Massol2017a}, as well as macro-ecological patterns including the speed of invasion waves, for example.

\subsection*{Lessons from existing theory}
In order to better define future research avenues we have discussed above what is known from the metapopulation literature. If we summarise our understanding of single species systems to a very abstract level, we can reduce dispersal theory to a single fundamental proposition and two fundamental principles: The proposition is that dispersal and its evolution are driven by the contrast between local (inclusive) fitness expectations and the fitness expectation elsewhere \citep[see e.g.,][]{Metz2001, Poethke2002}. The two general principles, to which most studies looking at more specific effects can be traced back to, are the following: Dispersal will be more frequent in systems where (1) the temporal variance in fitness expectations is large and (2) the spatial correlation in fitness expectations is low \citep[see e.g.][]{Hastings1983, Comins1980, McPeek1992}, i.e. where the spatio-temporal variance in fitness expectations is large. 

At the same level of abstraction nothing changes when we consider multi-species systems: Individuals should still emigrate whenever (inclusive) fitness expectations elsewhere are larger than those at the current site of residence and dispersal should occur more often whenever the spatio-temporal variation in fitness expectations is large. Whatever forms the ecological theatre for the evolutionary dispersal play, intra- or inter-specific competition, predation, parasitism or mutualism, does not matter in detail. At a lower level of abstraction, drivers may change in relative importance (e.g., kin competition) and additional drivers, such as co-dispersal, may directly impact costs and benefits of dispersal.

\subsection*{Directions for future research}
Since many points discussed above are largely speculative, they call for more investigation, both theoretically and empirically. In the following we provide a non-comprehensive list of potentially relevant questions and topics.

Going back to the most abstract level mentioned above, it is relevant to determine in general, what the most important drivers of fitness expectations for any species are (abiotic conditions, intra-specific or inter-specific interactions). These define the ecological theatre (selection pressures) for the evolutionary play. Moreover, it is relevant to ask which of these drivers are responsible for the largest temporal variance in fitness expectations and what the characteristic spatial domain of autocorrelation in this variance is. Further, with an increasing number of interactions, is spatio-temporal variance in fitness-expectations likely to increase or decrease? Are fitness expectations at all predictable in a world of highly complex and dynamic interactions?

From a theoretical point of view we have noted that co-dispersal is largely understudied. While some work exists at the interface between evolutionary epidemiology and dispersal evolution, as recently reviewed by \citep{Ziliosubmitted}, a comprehensive theoretical treatment is lacking. 

The same holds true for the joint evolution of dispersal with interaction traits, as well as considering higher-order interactions involved in dispersal (e.g. dispersal of parasites with complex life cycles and multiple successive host species). Future work could identify which ``parameters'' of interactions are important for dispersal (interaction signs, immediacy, effect on abundances, correlations with other traits, to name but a few). Dispersal syndromes with or without co-dispersers (importance of symbiosis for dispersal plasticity and correlation with other traits) may have to be more clearly included into theory. We would also like to highlight that the step from communities and food webs towards ecosystems has to be taken, e.g. by studying constraints on dispersal at the ecosystem level \citep[see][]{Massol2017}. Feedbacks will be interesting to understand, namely feedbacks between the evolution of dispersal in multiple species, but also feedbacks on local and regional community dynamics and realized interaction strengths, which can in turn modify dispersal evolution.

At the opposite end of the continuum of biological organisation levels, the genome, \citet{Debarre2014} show that epistasis and trait correlations may to some degree modulate responses, and the ``cost of complexity'' model suggests that modularity in genetic architecture may be a solution to simplify the problem of adaptation. This overall calls for putting the genetic architecture of traits, including dispersal, at the centre of theoretical and empirical investigations \citep{Saastamoinen2018, Deshpande2022, Fronhofer2023, Yamamichi2022}. Using a multilevel network modelling approach \citep{Melian2018} represents a promising way forward.

From an empirical point of view dispersal plasticity certainly requires more investigation. One could imagine, for instance, that a symbiont's dispersal strategy may be less plastic, being more constrained by the host dispersal plasticity, or could it be that symbiont dispersal plasticity is induced by host dispersal plasticity (when the host's environment changes, the same is true for the symbiont)? More generally, it would be relevant to understand what reliable cues, triggers or correlates of fitness-relevant drivers exist at the meta-food web level and how accurately such interspecific information can be perceived. Particularly, increasing the numbers and diversity of interactions may make every single one less fitness-relevant. If dispersal decisions are plastic, external cues may thus become less relevant and reliable in comparison to internal states like body mass/size, hunger level or health status, for example. 

\subsection*{Summary and conclusion}
In order to clear the way, at least conceptually, and advance our understanding of relevant drivers of dispersal evolution in species-diverse meta-systems, we here discussed whether theory built with single-species systems in mind holds at the multi-species level. Are effects exacerbated or averaged out? Do they change in relative importance?

We highlight that at the highest level of abstraction dispersal theory consists of a single fundamental proposition and two fundamental principles that presumably hold within and across species. At a lower level of abstraction, biotic interactions form the ecological theatre for the evolutionary dispersal play: Biotic interactions like trophic interactions mediate patterns of fitness expectations in space thereby reducing the realm of possible patterns and providing us with a modelling tool. While this implies a simple transposition of already known drivers to a multispecies context, some drivers may require some upscaling or more complex transpositions, such as kin effects and sexual reproduction. The latter provides interesting possibilities for linking micro- and macroevolutionary dynamics. Above, we have identified an important quantitative modulator of dispersal evolution: increased trait dimensionality which may impact the speed of evolution. Finally, an additional factor to consider is co-dispersal. The latter may lead to scale and selection pressure mismatches between hosts and symbionts. A large number of open questions and important potential consequences in both ecological and evolutionary terms calls for more investigation.

\section*{Author contributions}
All authors discussed the content and topic of the study during a workshop organized by E.A.F. F.M. and E.A.F. wrote a first draft of the manuscript and all authors commented on the draft.

\section*{Acknowledgements}
The idea of the study originated from a workshop on dispersal led by Emanuel A. Fronhofer supported by a grant from the Agence Nationale de la Recherche (No.: ANR-19-CE02-0015) to EAF. We thank all participants to the workshop for fruitful discussions. This is publication ISEM-YYYY-XXX of the Institut des Sciences de l'Evolution --- Montpellier, and BRCXXX of the Biodiversity Research Centre at UCLouvain. Nicolas Schtickzelle is Senior Research Associate of the F.R.S.-FNRS and acknowledges its financial support (project T.0211.19) as well as support from Action de Recherche Concertée (DIVERCE 18-23/095).

\end{document}